\documentclass[aps,pra,twocolumn,superscriptaddress,raggedbottom,showpacs,floatfix]{revtex4-1}
\usepackage{amsmath,amsfonts,amssymb,amsthm}
\usepackage{bm}
\usepackage{color}
\usepackage{graphicx,latexsym}
\usepackage{epstopdf}
\usepackage{comment}

\begin{document} 
\renewcommand{\vec}{\mathbf}
\renewcommand{\Re}{\mathop{\mathrm{Re}}\nolimits}
\renewcommand{\Im}{\mathop{\mathrm{Im}}\nolimits}

\title{Moving solitons in a one-dimensional fermionic superfluid}
\author{Dmitry K. Efimkin}
\affiliation{Joint Quantum Institute and Condensed Matter Theory Center, Department of Physics, University of Maryland, College Park, Maryland 20742-4111, USA}
\affiliation{School of Physics, Monash University, Melbourne, Victoria 3800, Australia}
\author{Victor Galitski}
\affiliation{Joint Quantum Institute and Condensed Matter Theory Center, Department of Physics, University of Maryland, College Park, Maryland 20742-4111, USA}
\affiliation{School of Physics, Monash University, Melbourne, Victoria 3800, Australia}

\begin{abstract}
A fully analytical theory of a traveling soliton in a one-dimensional  fermionic superfluid is developed within the framework of time-dependent self-consistent Bogoliubov-de Gennes equations, which are solved exactly in the Andreev approximation. The soliton manifests itself in a kink-like profile of the superconducting order parameter and hosts a pair of Andreev bound states in its core. They adjust to soliton's motion and play an important role in its stabilization. A phase jump across the soliton and its energy  decrease with soliton's velocity and vanish at the critical velocity, corresponding to the Landau criterion, where the soliton starts emitting quasiparticles and becomes unstable. The  ``inertial'' and ``gravitational''  masses of the soliton are calculated and the former is shown to be orders of magnitude larger than the latter. This results in a slow motion of the soliton in a harmonic trap, reminiscent to the observed behavior of a soliton-like texture in related experiments in cold fermion gases [T. Yefsah et al., Nature {\bf 499}, 426, (2013)]. Furthermore, we calculate the full non-linear dispersion relation of the soliton  and solve  the classical equations of motion in a trap. The strong non-linearity at high velocities gives rise to anharmonic oscillatory motion of the soliton. A careful analysis of this anharmonicity may provide a means to experimentally measure the non-linear soliton spectrum in superfluids.   
\end{abstract}
\maketitle

\section{Introduction.}
Solitons are fascinating non-linear phenomena that occur in a diverse array of classical and quantum systems (see, {\em {\em e.g.}}, Ref.~[\onlinecite{SolitonReviewGeneral}] and references therein). In particular, they are known to exist in quantum superfluids, and have been demonstrated experimentally in Bose-Einstein condensates (BECs) using various methods including phase imprinting ~\cite{PhaseImprinting0,PhaseImprinting1}, density engineering ~\cite{DensityEngeneering1,DensityEngeneering2}, and matter-wave interference ~\cite{MatterWaveInterference1} methods. A rich theoretical literature on solitons in BECs has also developed~\cite{SolitonReviewBEC1, PitaevskiiStringari} and it includes both numerical and analytical solutions of Gross-Pitaevskii equations in excellent agreement with both each other and experiment. 

Fermionic superfluids also support solitons - a phase jump in the order parameter field. These are more interesting  and complicated objects than ``Gross-Pitaevskii solitons,'' because they can host and carry localized fermionic excitations - Andreev bound states (ABS).  Consequently, a description of these non-linear phase excitations is more complicated: there exists no closed equation for the bosonic order parameter field and to include fermionic degrees of freedom is essential. At the technical level, one has to solve two-component Bogoliubov-de Gennes (BdG) equations supplemented with a non-linear self-consistency constraint.  This class of problem in one dimension has been studied extensively in the context of  the Gross-Neveu model  of quantum field theory ~\cite{QFTClassics1,QFTClassics2, Dunne2, Nitta1, Nitta2, Thies1, Thies2, Thies3}, organic polymers ~\cite{TakayamaLinLiuMaki, KivelsonLeeLinLiuPeschenYu, CDWReview, Brazovskii1, Brazovskii2}, and mesoscopic superconductivity ~\cite{Machida1, Machida2, Yakovenko1, Yakovenko2} (see also Ref.~[\onlinecite{PhaseSlips}] for the Eilenberger approach to a related problem of phase slips in one-dimensional superconductors). Using remarkable connections to inverse scattering method and supersymmetric quantum mechanics, exact analytical solutions were found to describe {\em static} soliton textures.    

More recently, numerical analyses of \emph{static} and {\em moving} solitons in neutral fermionic superfluids within the crossover from BEC to BCS (Bardeen-Cooper-Schrieffer) regimes were developed~\cite{SolitonBECBCS1,SolitonBECBCS2,SolitonBECBCS3,SolitonBECBCS4,SolitonBECBCS5,SolitonBECBCS6, Sacha1, Sasha2}. On the experimental side, the Zwierlein group at MIT reported in 2013 an observation of an oscillating solitonic vortex (that is actually a three-dimensional vortex-like texture, which tends to the soliton in the limit of the true one-dimensional confinement) in a strongly-interacting fermionic superfluid in an elongated trap~\cite{Zwierlein1,Zwierlein2} (see also \cite{Levin} for a discussion of  stability of soliton-like textures). These developments, along with potential connections to Majorana fermions  (which may be carried by solitons in one-dimensional topological superfluids \cite{SolitonMajorana,TopologicalQuantumCompuatation}), make the problem of fundamental understanding of soliton dynamics in one-dimensional paired Fermi systems of significant importance and interest.

Here, we develop an {\em analytic theory of a traveling soliton} in a one-dimensional  paired superfluid in the weak coupling BCS regime. We show that the time-dependent BdG equations are exactly solvable in the Andreev approximation to describe a uniformly-moving solitary wave of the BCS order parameter and  derive a  dependence of the soliton's energy and phase discontinuity across it on its velocity. The two latter quantities are shown to decrease monotonically with velocity and vanish at  the Landau critical velocity.  It is also shown that the ABS, carried by the soliton, adjust to its motion and play an important role in its stabilization.  The ``inertial'' and ``gravitational''  masses of the soliton are calculated and the former is shown to be orders of magnitude larger than the latter. This results in a slow motion of the soliton in a harmonic trap, reminiscent to what has been observed in the
relevant experiment ~\cite{Zwierlein1,Zwierlein2}. At high velocities, the non-linearity of soliton spectrum becomes essential and it leads to anharmonic oscillations, expressed in terms of elliptic functions. 

The rest of the paper is organized as follows. In Sec.~II, the time-dependent Bogoliubov-de Gennes equations are introduced. In Sec.~III, we construct their self-consistent solution, which describes a moving solitary wave. Sec.~IV is devoted to soliton energetics. In Sec.~V, we consider soliton dynamics in a trap, calculate soliton's effective masses, and solve the classical equations of motion including the full non-linear spectrum.  We conclude in Sec.~VI.

\section{Time-dependent mean-field theory}
We start with the BCS model for a one-dimensional uniform superfluid, written in the Heisenberg representation
\begin{equation}
H=\int dx \left[\sum_{\alpha} \Psi^+_\alpha \epsilon(\hat{p}_x) \Psi_\alpha- V \Psi^+_\uparrow \Psi^+_\downarrow \Psi_\downarrow \Psi_\uparrow \right].
\label{theH}
\end{equation}
Here $\Psi_{\alpha}\equiv\Psi_{\alpha}(x,t)$ ($\Psi_{\alpha}^+\equiv\Psi_{\alpha}^+ (x,t)$) is the annihilation (creation) Heisenberg operator for fermions, which can be written in the Nambu representation $\Psi=\{\Psi_{\uparrow}, \Psi^+_{\downarrow} \}^T$; $\epsilon(\hat{p})=(\hat{p}_x^2-p_\mathrm{F}^2)/2m$ is the kinetic energy of fermions; $V$  and  $\nu_\mathrm{F}$ are the attractive interaction and density of states on the Fermi level, leading to the dimensionless coupling constant $\lambda=V \nu_\mathrm{F}\ll 1$, which is a small parameter in the weak coupling BCS regime. The operators satisfy the equation of motion, $i \hbar \partial_t \Psi=[H,\Psi]$, which in \emph{time-dependent} mean-field approach~\cite{Comment1} with the order parameter, $\Delta(x,t)=-V \langle \Psi_\downarrow (x,t)\Psi_\uparrow (x,t)\rangle$, reduces to 
\begin{equation}
\label{BdGH}
i \hbar \partial_t \Psi(x,t)=\left(\begin{array}{cc} \epsilon(\hat{p}_x) & \Delta (x,t) \\ \Delta^* (x,t) & -\epsilon(\hat{p}_x)  \end{array} \right)\Psi(x,t).
\end{equation}
The matrix operator in the above equation is the time-dependent BdG Hamiltonian. We seek a uniformly-moving solution, where the order parameter and field operators are functions of  the single variable, $z=x+v_\mathrm{s}t$. In the weak coupling regime, the semiclassical (Andreev) approximation~\cite{Andreev}, which treats separately the left- ($\alpha=-1$) and right-moving ($\alpha=+1$) fermions can be employed. We present the field operator in the form $\Psi (x,t)=\sum_{\alpha n} \psi^\alpha_{n}(z) b^\alpha_{n} \exp[i( \alpha p_\mathrm{F} z - \epsilon^\alpha_{n} t)/\hbar]$, where the sum is over time-dependent Bogoliubov quasiparticle's states, described by the operators $b^\alpha_{n}$, with the energies $\epsilon^\alpha_{n}$ and wave functions $\psi^\alpha_{n}(z)=\{u^\alpha_{ n} (z), v^\alpha_{n}(z)\}^T$. The Ansatz for $\Psi (x,t)$ satisfies the equation of motion, Eq.~(\ref{BdGH}), if the Bogoliubov states satisfy $K_\mathrm{BdG}^\alpha(z) \psi^\alpha_n(z)=\epsilon^\alpha_n \psi^\alpha_n(z)$ with the effective Hamiltonian
\begin{equation}
\label{BdGKSC}
K_\mathrm{BdG}^\alpha=
\left(\begin{array}{cc} \alpha v_\mathrm{F} \hat{p}_z+\alpha v_\mathrm{s} p_\mathrm{F}    & \Delta (z) \\ \Delta^* (z) &-\alpha v_\mathrm{F} \hat{p}_z +\alpha v_\mathrm{s} p_\mathrm{F}   \end{array} \right),  
\end{equation}
which does not have an explicit time dependence and corresponds to the frame of reference moving together with the soliton. It differs from the time-dependent  Hamiltonian in the original laboratory frame
\begin{equation}
\label{BdGHSC}
H_\mathrm{BdG}^\alpha=
\left(\begin{array}{cc} \alpha v_\mathrm{F} \hat{p}_x   & \Delta (x+v_\mathrm{s}t) \\ \Delta^* (x+v_\mathrm{s}t) &-\alpha v_\mathrm{F} \hat{p}_x \end{array} \right)  
\end{equation}
by the energy shift $\delta \epsilon^\alpha=\alpha v_\mathrm{s} p_\mathrm{F}$. As a result, in this co-moving frame, we assume Bogoliubov quasiparticles to be in thermal equilibrium and the self-consistent equation  for the order parameter becomes,
\begin{equation}\label{SelfConsistent}
\Delta(z)=-V \sum_{\alpha n} u^\alpha_{n}(z) [v^\alpha_{n}(z)]^* n_\mathrm{F} (\epsilon^\alpha_{n}),
\end{equation}
where $n_\mathrm{F} (\epsilon^\alpha_{n})$ is the thermal Fermi-Dirac distribution function. The equation has a uniform solution, corresponding to the BCS superfluid state with the uniform order parameter, $\Delta_0\sim E_\mathrm{F} \exp[-1/\lambda]$, but it also has  nontrivial solitonic solutions. 

Note that we have reduced the \emph{time-dependent} many-body problem to a \emph{ time-independent} one with the energy shift $\delta \epsilon^\alpha$ of Bogoliubov quasiparticle's energies. The shift does not change the general structure of the BdG Hamiltonian and enables us to use the machinery developed in the context of static solitons. Nevertheless, since energy shifts for right- and left- Fermi points have opposite signs, they modify the energetics of the solitonic solutions in a non-trivial fashion and are essential for the following.  

\begin{figure}
\label{Fig1}
\begin{center}
\includegraphics[width=8.5 cm]{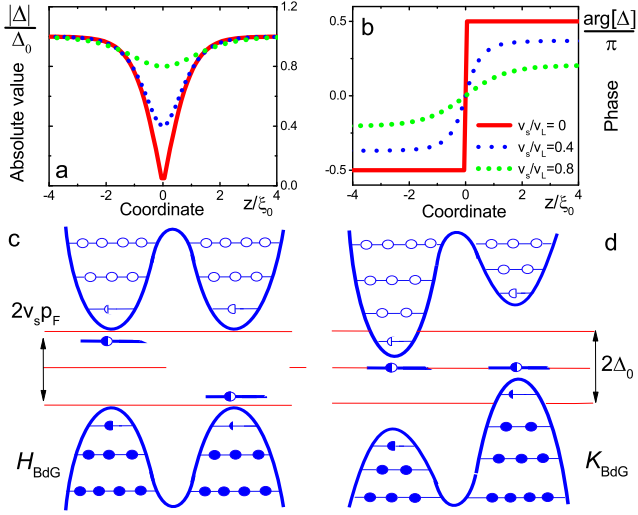}
\caption{(Color online)  Figs.~1a and 1b show spatial profiles of the absolute value and phase of the order parameter, respectively,  plotted for different soliton velocity, $v_\mathrm{s}$. Figs.~1c and 1d represent the energy spectra of the BdG Hamiltonians in the laboratory frame  ($H_\mathrm{BdG}$)  and co-moving frame ($K_\mathrm{BdG}$)  respectively. Filled and empty circles denote occupied and empty Bogoliubov states, accordingly. Incomplete circles correspond to a decreasing number of states in the continuous Bogoliubov bands  due to ABS splitting from them. In a solitonic state, the energies of ABS corresponding to $K_\mathrm{BdG}$, are exactly zero, while their energies corresponding to  $H_\mathrm{BdG}$, are split and shifted away from zero by $\pm v_\mathrm{s}p_\mathrm{F}$.}    
\end{center}
\end{figure}
\section{Solitonic solutions}
In the Andreev approximation, the problem [see, Eqs.~(\ref{BdGKSC}) and~(\ref{SelfConsistent})] maps to the Gross-Neveu model, for which self-consistent solitonic solutions can be found exactly~\cite{QFTClassics1,QFTClassics2}. Particularly, it was shown that both BdG equations~(\ref{BdGKSC}) and Eq.~(\ref{SelfConsistent})  are simultaneously satisfied,  if the order parameter yields a reflectionless potential for Bogoliubov quasiparticles. In that case, the BdG equations reduce to a pair of supersymmetric  Schr{\"o}dinger equations, see Eq.~(\ref{SSS}) below, which can be solved exactly. A family of reflectionless potentials, corresponding to a single localized soliton, can be parameterized by a phase jump, $2\phi$,  across it as follows 
\begin{equation}
\label{SolitonicProfile}
\Delta(z)=\Delta_0 \left\{\cos(\phi) + \bm{i} \sin(\phi) \tanh [ \sin (\phi) \cdot z_\xi] \right\}.
\end{equation}
Here $z_\xi=z/\xi_0$, where $\xi_0=\hbar v_\mathrm{F}/\Delta_0$ is the coherence length.  The spatial dependencies of the order parameter's phase and modulus are presented in Fig.~1. At $2\phi=0$ the solitonic texture vanishes and the order parameter profile becomes uniform. Introducing $f^\alpha_{\pm}(z)=u^{\alpha}(z)\pm v^\alpha(z)$, the BdG equations can be reduced to a pair of equations
\begin{equation}
\label{SSS}
\left[-\hbar^2 v_\mathrm{F}^2 \partial_z^2+|\Delta(z)|^2 \pm \alpha \hbar v_\mathrm{F} \frac{\partial\Delta_2 (z)}{dz}\right] f^\alpha_{\pm}=\epsilon^2 f^\alpha_{\pm},
\end{equation}
that have supersymmetric (SUSY) structure (See Ref.~[\onlinecite{SuSy}] for a review). Particularly, they can be presented as $H_\pm^\alpha f_\pm^\alpha =E f_\pm^\alpha$ with the effective energy $E=\epsilon^2-\Delta_1^2$ and Hamiltonians $H_\pm^\alpha=A_\mp^\alpha A_\pm^\alpha$, which are a product of the ladder operators $A_{\pm}^\alpha=-i \hbar v_\mathrm{F}\partial_z\pm \alpha  i \Delta_2(z)$.  Here, the imaginary part of the order parameter $\Delta_2(z)$ plays the role of the superpotential $W(z)$~[\onlinecite{SuSy}]. The presence of a kink  in its spatial dependence, where the order parameter changes sharply from $-\Delta_0 \sin(\phi)$ to $\Delta_0 \sin(\phi)$, guaranties the existence of a localized solution for one of these equations (\ref{SSS}). Using the explicit profile of the order parameter (\ref{SolitonicProfile}), we cast the BdG equations into the following form
\begin{equation}
\begin{split}
\left[\hbar^2 v_\mathrm{F}^2 \partial_z^2-\Delta_0^2+\epsilon^2 \right] f^{\alpha}_{\alpha}=0, \\
\left[\hbar^2 v_\mathrm{F}^2 \partial_z^2-\Delta_0^2\left\{1-\frac{2\sin^2(\phi)}{\cosh^2[\sin(\phi) z_\xi]} \right\} +\epsilon^2\right] f^{\alpha}_{\bar{\alpha}}=0.
\end{split}
\end{equation}
The equation for $f^\alpha_{\alpha}$ is trivial and contains only a continuous spectrum with plane-wave solutions, while the equation for $f^\alpha_{\bar{\alpha}}$ has both the continuous states and an extra bound state. The continuous solutions have energy, $\epsilon_{\gamma k}=\gamma \sqrt{(\hbar v_\mathrm{F} k)^2 + \Delta_0^2}\equiv \gamma \epsilon_{k}$, where $\gamma=\pm1$ corresponds to the Bogoliubov particles and holes, and are given by
\begin{equation}
\begin{split}
\label{ContinousStates}
u^\alpha_{\gamma k}(z)=\sqrt{\frac{\epsilon_{\gamma k}+ \alpha \Delta_1}{4 L  \epsilon_{\gamma k} }} \left[1 +  \alpha \frac{ \hbar v_\mathrm{F}k+\bm{i} \Delta_2(z)}{\epsilon_{\gamma k}+ \alpha \Delta_1} \right] e^{\bm{i} k z},\\
v^\alpha_{\gamma k}(z)=\sqrt{\frac{\epsilon_{\gamma k}+ \alpha \Delta_1}{4 L  \epsilon_{\gamma k} }}\left[\alpha -  \frac{ \hbar v_\mathrm{F}k+\bm{i} \Delta_2(z)}{\epsilon_{\gamma k}+ \alpha \Delta_1} \right]e^{\bm{i} k z}.
\end{split}
\end{equation}
Andreev bound states, localized on the soliton, have the energy $\epsilon	^\alpha_{\mathrm{ABS}}=-\alpha \Delta_0 \cos \phi$ and are described by the following wave functions
\begin{equation}
\label{AndreevBoundStates}
\psi^\alpha_{\mathrm{ABS}}(z)=\frac{1}{2} \sqrt{\frac{\sin(\phi)}{\xi_0}} \frac{1}{\cosh[\sin(\phi)z_\xi]}\left(\begin{array}{cc} 1 \\ -\alpha \end{array}\right).
\end{equation}
The energies of ABSs are sensitive to the phase jump across the soliton, while the dispersion law of Bogoliubov quasiparticles remains unchanged in the presence of the soliton compared to the uniform BCS state. However, the solitonic texture modifies the density of states of the Bogoliubov particles and holes. Indeed, for the sake of qualitative argument, consider an adiabatic insertion of a soliton from the uniform state. In this adiabatic process, the Andreev bound states are split from the continuous particle and hole bands, but the total number of fermionic states is conserved. Therefore, the continuous bands for each Fermi point have one state less compared to the uniform superfluid. 

The presence of a soliton distorts boundary conditions, which can not longer be considered as simple periodic, and modifies the momentum quantization. Indeed, while all \emph{local} physical observables [{\em e.g.}, the fermion current $j(z)$, density $\rho(z)$, {\em etc.}] are periodic functions of the coordinate in a closed system [$j(z+L/2)=j(z-L/2)$, $\rho(z+L/2)=\rho(z-L/2)$, {\em etc.}], the order parameter is not periodic, because it has a \emph{global} phase discontinuity across the soliton, and $\Delta(z+L/2)=\Delta(z-L/2)e^{2\bm{i} \phi}$. Here $L$ is the system length. We have generalized the periodic boundary conditions for a system with a soliton (see Appendix A for their detailed derivation), and they are given by 
\begin{equation}
\psi^\alpha_{\gamma k}(z+L/2)=\left[\cos(\phi)+{\bm i} \sin(\phi) \sigma_z\right]\psi^\alpha_{\gamma k}(z-L/2).
\end{equation}
They reduce to simple periodic boundary conditions, $\psi^\alpha_{\gamma k}(z+L/2)=\psi^\alpha_{\gamma k}(z-L/2)$, if the phase jump $\phi=0$, when the soliton vanishes and the order parameter becomes uniform. Using the explicit form of the wave functions (\ref{ContinousStates}), we obtain the  quantization condition for quasiparticle's momentum $k_n L+\theta^\alpha_{\gamma}(k_n)=2 \pi n$, where $n$ is integer and
\begin{equation}
\theta^\alpha_{\gamma}(k)=\mathrm{arg}\left[\epsilon_{k} \cos(\phi) + \alpha\gamma \Delta_0 - \bm{i}  \alpha \gamma \hbar  v_\mathrm{F} k \sin(\phi) \right]
\label{pshift}
\end{equation}
is a phase shift (the calculations are presented in Appendix B).  Using these phase shifts, we find the number of states $N^\alpha_\gamma $, split from the  continuous bands, as follows~\cite{TakayamaLinLiuMaki}
\begin{equation}
N^\alpha_\gamma= -\int_{-\infty}^\infty \frac{dk}{2\pi} \frac{d \theta^\alpha_\gamma}{dk}=\frac{1}{2}-\alpha \gamma\left(\frac{1}{2}-\frac{\phi}{\pi}\right).
\end{equation}
leading to $N^\alpha_\alpha=\phi/\pi$ and $N^\alpha_{\bar{\alpha}}=(\pi-\phi)/\pi$. Since there is the only one ABS per a Fermi point, the sum of these numbers is $N^\alpha_++N^\alpha_-=1$, which confirms the physical picture of ABS splitting off of the Bogoliubov bands. The total number of states split from the valence and conduction bands is also an integer: $N^+_-+N^-_-=1$, and $N^+_++N^-_+=1$. 

The energies of the continuous states and ABS  in the co-moving frame are shifted by $\delta \epsilon^\alpha=\alpha v_{\mathrm{s}}p_\mathrm{F}$. For the continuous spectrum this shift is unimportant as long as $v_{\mathrm{s}}\le v_\mathrm{L}$, where $v_\mathrm{L}=\Delta/p_\mathrm{F}$ is the critical velocity within the Landau criterion. At $v=v_\mathrm{L}$, the continuous bands touch the zero energy level and soliton can lower its energy by emitting Bogoliubov excitations and becomes unstable. For localized states, the energy shift is crucial since it governs both the energy and occupation of these states. 
\begin{figure}
\label{Fig2}
\begin{center}
\includegraphics[width=8.5 cm]{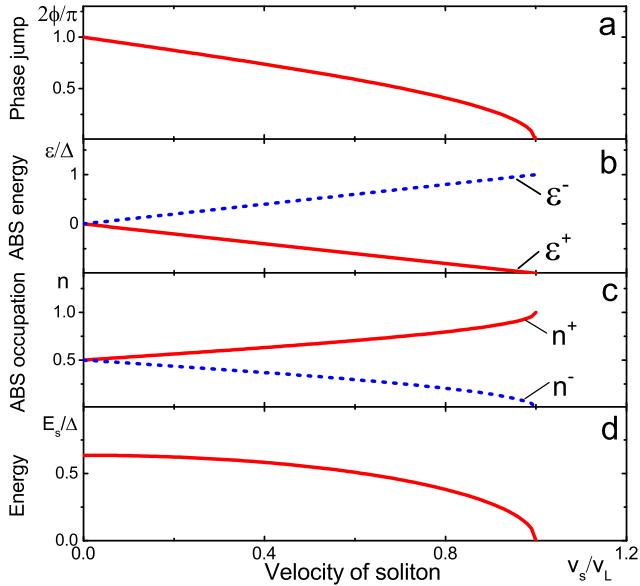}
\caption{(Color online) Plotted are the velocity dependence of: (a)~the phase jump across the soliton, $2\phi_\mathrm{s}$; (b)~energies of the ABS, $\epsilon^\alpha_{\mathrm{ABS,s}}$, localized on the soliton; (c)~their occupation numbers ,$n^{\alpha}_{\mathrm{ABS,s}}$; (d) energy of the moving soliton in the laboratory frame, $E_\mathrm{s}$.}    
\end{center}
\end{figure}

So far the phase jump across a soliton, $2\phi$, has been treated as an independent parameter characterizing the shape of the order parameter within the  family of reflectionless potentials, given be Eq.~(\ref{SolitonicProfile}). However, its value is fixed by the self-consistent equation for the order parameter (\ref{SelfConsistent}), which we have not take into account yet. Due to the self-consistency constraint, the phase jump becomes dependent on the soliton velocity $v_\mathrm{s}$.   Using semiclassical wave functions (\ref{ContinousStates}) and (\ref{AndreevBoundStates}), the self-consistent equation for order parameter (\ref{SelfConsistent}) can be rewritten as  
\begin{equation}
\label{SelfConsistentEquation2}
\begin{split}
\Delta(z)= \frac{V \Delta_0}{4\hbar v_\mathrm{F}} \delta n \frac{\sin \phi}{\cosh^2[\sin(\phi) z_\xi ]} + \\ + V \int \frac{dk}{2\pi} \frac{\Delta(z)}{\epsilon_k} 
- \frac{V \Delta_0}{4\hbar v_\mathrm{F}} \frac{\pi-2 \phi}{\pi} \frac{\sin \phi}{\cosh^2[\sin(\phi) z_\xi]},
\end{split}
\end{equation}
where $\delta n=n_\mathrm{+}-n_\mathrm{-}=n_\mathrm{F} [v_\mathrm{s} p_\mathrm{F}-\Delta_0 \cos(\phi)] - n_\mathrm{F} [-v_\mathrm{s} p_\mathrm{F}+\Delta_0 \cos(\phi)]$ is a difference between the occupation numbers of the ABS, which are influenced by the soliton's motion. The latter two terms originate from the continuous Bogoliubov states, and for them we can set  the temperature to zero. However, the zero-temperature limit for ABS is delicate, because it implies $T\ll |v_\mathrm{s} p_\mathrm{F}-\Delta_0 \cos(\phi)|$, which can not hold when the corresponding energies vanish, while Fermi distribution functions in the zero temperature limit become non-monotonous.  The self-consistent equation  (\ref{SelfConsistentEquation2}) is satisfied if 
\begin{equation}
\label{SelfConsistentEquation3}
\sin (\phi) \left[ \pi-2\phi - \pi \delta n \right]=0.
\end{equation}
This equation has the trivial solution $2\phi=0$, which corresponds to a uniform BCS state with no solitons. It also has a \emph{single} nontrivial solution, corresponding to a traveling soliton with the phase jump across it, which in the zero-temperature limit takes the simple form
\begin{equation}
\label{PhaseProfile}
2 \phi_\mathrm{s} =2 \arccos\left(\frac{v_\mathrm{s}}{v_\mathrm{L}}\right).
\end{equation}
Note that the  energies of ABSs are zero in the co-moving frame, while in the laboratory frame they are split in energy by $\epsilon^\alpha_{\mathrm{ABS,s}}=-\alpha v_\mathrm{s}p_\mathrm{F}$. The occupation numbers of ABS adjust to  soliton's motion and are not equal. The occupation numbers can be calculated from Eq.~(\ref{SelfConsistentEquation3}) as follows
\begin{equation}
n^{\alpha}_\mathrm{ABS,s}=\frac{1}{2} +\alpha \left[\frac{1}{2} - \frac{\phi_\mathrm{s}(v_\mathrm{s})}{\pi}\right].
\end{equation}	
The dependencies of phase jump across the soliton, energies and occupations of ABSs on  velocity $v_\mathrm{s}$ are presented in Figs.~2a\,--\,2c. The soliton at rest has a phase jump of $2\phi_\mathrm{s}=\pi$ across it, while ABSs have zero energies and they are equally occupied  $n^{\alpha}_\mathrm{ABS,s}=1/2$, as have been previously derived~\cite{TakayamaLinLiuMaki,Brazovskii1}. The phase jump decreases with velocity $v_\mathrm{s}$ until the critical one $v_\mathrm{L}$ is reached. The splitting of ABSs energies $2 v_\mathrm{s} p_\mathrm{F}$ and difference between their occupations $\delta n_\mathrm{s}=1-2 \phi_\mathrm{s}/\pi$ gradually increase with the soliton's velocity. 

The total occupation of the ABS is equal to one (i.e., $n^+_{\mathrm{ABS,s}}+n^-_{\mathrm{ABS,s}}=1$), which coincides with the number of states split off of the lower Bogoliubov band (i.e., $N^+_-+N^-_-=1$). It means that within the Andreev approximation there is neither a deficit, nor an excess of fermionic matter in the soliton core compared to the uniform state: $\delta N_\mathrm{s}=0$.
It should be noted, that in the local density approximation, the deficit (or excess) of fermions determines the interaction strength of the soliton with a trap potential, confining the superfluid, and its sign is crucial for soliton dynamics. Below, we show that more general thermodynamic arguments give a small but finite value for $|\delta N_\mathrm{s}|\sim \Delta_0/\lambda E_\mathrm{F}$ [see Eq.~(\ref{ParticleDefficit})], which can be both positive, and negative, depending on the sign of the energy derivative of the density of states, which in turn is determined by the (true) dimensionality of the system and geometry of the Fermi surface.

\section{Soliton energetics} In equilibrium, the self-consistency constraint corresponds to an extremum or a saddle point of the free energy of the system (energy in the zero-temperature limit). Our time-dependent approach involves a mapping of the time-dependent Hamiltonian in the laboratory frame (\ref{BdGHSC}) on a time-independent model (\ref{BdGKSC}) with a ``distorted'' BdG Hamiltonian $K_\mathrm{BdG}$, with the velocity of the soliton, $v_\mathrm{s}$, playing the role of an external parameter. The corresponding energy, $E^\mathrm{K}(\phi,v_\mathrm{s})$,  in the co-moving frame achieves an extremum as a function of $\phi$, corresponding to the solution (\ref{PhaseProfile}).  However, the actual energy of the solitonic state in the laboratory frame, $E^\mathrm{H}(\phi,v_\mathrm{s})$, differs from $E^\mathrm{K}(\phi,v_\mathrm{s})$ as discussed below.
\begin{figure}
\label{Fig3}
\begin{center}
\includegraphics[width=8.5 cm]{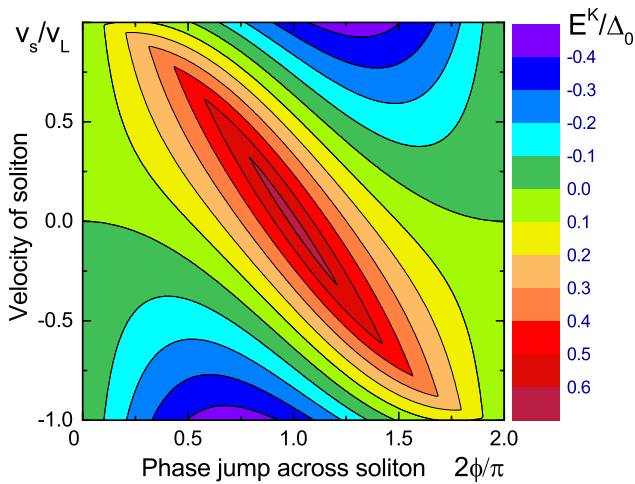}
\caption{(Color online) Shown is the dependence of the energy, $E^\mathrm{K}$, on the phase discontinuity across the soliton, $2\phi$, and its velocity $v_\mathrm{s}$. The dependence has a clear maximum, corresponding to the relation (\ref{PhaseProfile}), which holds when the BdG equations and the self-consistency equation are satisfied simultaneously.}    
\end{center}
\end{figure}

The difference between $E^\mathrm{K}(\phi,v_\mathrm{s})$ in the solitonic state and one in uniform BCS state can be presented as the sum $E^\mathrm{K}=E_\mathrm{\Delta}+E_{\mathrm{c}}^\mathrm{K}+E_{\mathrm{ABS}}^\mathrm{K}$, where $E_\mathrm{\Delta}$ comes directly from the non-uniformity of the order parameter
\begin{equation}
E_\mathrm{\Delta}= \frac{1}{V} \int dz \left[ |\Delta(z)|^2-\Delta_0^2 \right].
\end{equation}
The contribution $E_{\mathrm{c}}^\mathrm{K}$ originates from filled continuous Bogoliubov states and can be calculated using Eq.~(\ref{pshift}) as follows~\cite{TakayamaLinLiuMaki}
\begin{equation}
E_\mathrm{c}^\mathrm{K}=\sum_{\alpha}\left[{N^{\alpha}_{-}} \Delta_0 + \sum_{k} \theta^{\alpha}_{-} \frac{\partial \epsilon_{k}}{\partial k}\right] - v_\mathrm{s}p_\mathrm{F} (N^+_- - N^-_-),
\end{equation}
with the last term here coming from the asymmetry between the states split from the continuum at the right and left Fermi points. Finally, the contribution, $E_{\mathrm{ABS}}^\mathrm{K}$, originates from the ABS and is given by
\begin{equation}
E_{\mathrm{ABS}}^\mathrm{K}=[v_\mathrm{s}p_\mathrm{F} - \Delta_0 \cos(\phi)] \delta n.
\end{equation} Putting all the three terms together (Detailed calculations are presented in Appendix C), we arrive at the following soliton energy in the co-moving frame
\begin{equation}
\begin{split}
E^\mathrm{K}(\phi,v_\mathrm{s})=\frac{2 \Delta_0}{\pi} \left[ \sin(\phi)+\left( \frac{\pi}{2} - \phi\right)\cos(\phi)\right]- \\ - v_\mathrm{s} p_\mathrm{F} \left(1-\frac{2 \phi}{\pi}\right)
- |v_\mathrm{s}p_\mathrm{F} - \Delta_0 \cos(\phi)|.
\end{split}
\end{equation}
The dependence of the energy $E^\mathrm{K}(\phi,v_\mathrm{s})$ on the phase jump $2\phi$ and the velocity $v_\mathrm{s}$ is presented in Fig.~3. For a soliton at rest, the energy has a clear maximum at $2\phi=\pi$. At a finite velocity, the energy maximum shifts and follows the curve corresponding to Eq.~(\ref{PhaseProfile}). This however does not imply that the corresponding solution is unstable and/or unphysical.  If we fix a phase jump across the soliton, which is a global constraint, the  solution found self-consistently from the BdG equations becomes a minimum of the corresponding energy functional~\cite{TakayamaLinLiuMaki} ({\em {\em e.g.}}, distorting the shape of the solitary wave would always increase the system's energy, as long as global boundary conditions are preserved). 
This means that the soliton is stable against local perturbations, which was confirmed in numerical simulations of the BdG equations~\cite{SolitonBECBCS3,SolitonBECBCS4,SolitonBECBCS5}. Interestingly, at a finite velocity, there appear additional local minima of $E^\mathrm{K}(\phi,v_\mathrm{s})$,  gradually emerging from the trivial solutions $2\phi=0, 2 \pi$ (see, Fig.~3). However, they do not satisfy the self-consistency constraint (\ref{SelfConsistentEquation2}), and hence are locally unstable.

The energy of the  system in the laboratory frame, $E^\mathrm{H}(\phi,v_\mathrm{s})$, follows from Hamiltonian~(\ref{BdGHSC}), can be calculated in the same manner as above (The calculations are presented in Appendix C) and is given by 
\begin{equation}
E_\mathrm{s}=E^\mathrm{H}(\phi_\mathrm{s}(v_\mathrm{s}),v_\mathrm{s})=\frac{2 \Delta_0}{\pi}\sqrt{1-\left(\frac{v_\mathrm{s}}{v_\mathrm{L}}\right)^2}.
\label{Lorentz}
\end{equation}
The energy of the soliton at rest is $E_\mathrm{s}(0)=2\Delta_0/\pi$. It gradually decreases with the velocity $v_\mathrm{s}$ and vanishes at the critical velocity $v_\mathrm{L}$, as presented in Fig.~2d.

\section{Soliton dynamics in a trap}
For a superfluid in a trap, the confining potential makes the soliton energy position-dependent and drives its motion. In the local density approximation, the chemical potential of fermions is $E_\mathrm{F}(x) = E_\mathrm{F}- U(x)$, where   $U(x)=m \omega^2 x^2/2$  is a harmonic trapping potential with frequency, $\omega$. The energy of a soliton with velocity $v_\mathrm{s}$ and coordinate $x_\mathrm{s}$ at $v_\mathrm{s}\ll v_\mathrm{L} $ and $U(x_\mathrm{s})\ll E_\mathrm{F}$ can be approximated as follows
\begin{equation}
E_\mathrm{s}(v_\mathrm{s},x_\mathrm{s})=\frac{2 \Delta_0}{\pi} + \frac{m_\mathrm{s}^\mathrm{i} v_\mathrm{s}^2}{2} + \frac{ m_\mathrm{s}^\mathrm{g} \omega^2 x_\mathrm{s}^2}{2},
\end{equation}
where $m_\mathrm{s}^\mathrm{i}$ and  $m_\mathrm{s}^\mathrm{g}$ are the ``inertial'' and ``gravitational'' masses, which define kinetic and potential energy of the soliton in the trap, and are given by   
\begin{equation}
\label{Masses}
m_\mathrm{s}^\mathrm{i}=-\frac{4 m}{\pi} \frac{E_\mathrm{F}}{\Delta_0}; \quad \quad \quad m_\mathrm{s}^\mathrm{g}=-\frac{2m}{\pi}\frac{\partial \Delta_0}{\partial  E_\mathrm{F}}. 
\end{equation}
The inertial mass of the soliton is always negative and is considerably larger than a single fermion's mass $m$. The negative sign of the mass implies that any dissipation (which can be introduced as $\dot{E}_\mathrm{s}=-\Gamma_\mathrm{s}|m_\mathrm{s}^\mathrm{i}| v_\mathrm{s}^2$ with $\Gamma_\mathrm{s}$ being a friction coefficient) would accelerate the soliton until it achieves the critical velocity and vanishes. The fermionic degrees of freedom (both the continuous states and ABSs) can play the role of a bath and lead to dissipation with $\Gamma_\mathrm{s}\sim  \Delta_0/\hbar\times \exp[-\Delta_0/T]$ \cite{Samokhin}. The dissipation is exponentially small at low temperatures $T\ll\Delta_0$ and can lead to a macroscopically large soliton life-time. 

In contrast to the inertial mass, the sign of the gravitational mass can be both positive and negative, depending on an energy dependence of the fermionic density of states $\nu_\mathrm{F}$ on the Fermi level, which determines the derivative $\partial \Delta_0/\partial E_\mathrm{F}\approx\ \Delta_0/\lambda^2 \times  \partial\lambda/\partial E_\mathrm{F} $ in Eq.~(\ref{Masses}). Particularly, in a truly one-dimensional fermionic superfluid (here we ignore the conceptual questions related to the possibility of superconductivity in such systems), the density of states decreases with energy $\partial \nu_\mathrm{F}/\partial E_\mathrm{F}=-\nu_\mathrm{F}/2E_\mathrm{F}$, which leads to a positive gravitational mass $m_\mathrm{s}^\mathrm{g}\approx m \Delta_0/\lambda \pi E_\mathrm{F}$. Note that the latter is considerably smaller than the mass of a single fermion $m$. According to the equation of motion for a soliton $\ddot{x}_\mathrm{s}-\Gamma_\mathrm{s}\dot{x}  -\omega_\mathrm{s}^2 x_\mathrm{s}=0$, it is accelerated away from the trap center with the rate  
\begin{equation}
\label{SolitonFrequency}
\omega_\mathrm{s}=\omega \sqrt{\left|\frac{m_\mathrm{s}^\mathrm{g}}{m_\mathrm{s}^\mathrm{i}}\right|}\approx\frac{\omega\Delta_0}{2 \sqrt{\lambda} E_\mathrm{F}}.
\end{equation}
In the more realistic and experimentally-relevant case of a quasi-one-dimensional fermionic superfluid with a circular Fermi surface (including a three-dimensional condensate in an elongated trap, such as studied in experiment~\cite{Zwierlein1}), the density of states increases with the energy $\partial \nu_\mathrm{F}/\partial E_\mathrm{F}=\nu_\mathrm{F}/2E_\mathrm{F}$ and the gravitational mass is negative $m_\mathrm{s}^\mathrm{g}\approx-m \Delta_0/\lambda \pi E_\mathrm{F}$.  Note that it is also considerably smaller than the mass of a single fermion, $m$. The equation of motion yields  $\ddot{x}_\mathrm{s}-\Gamma_\mathrm{s} \dot{x}_\mathrm{s} + \omega_\mathrm{s}^2 x_\mathrm{s}=0$, where $\omega_\mathrm{s}$, introduced in Eq.~(\ref{SolitonFrequency}), plays the role of an oscillation frequency of the soliton. Due to  dissipation, the soliton oscillates with an increasing amplitude, until it achieves the critical velocity, $v_\mathrm{L}$. A similar picture was observed for solitonic vortices in Refs.~[\onlinecite{Zwierlein1,Zwierlein2}].

The gravitational mass of the soliton, $m_\mathrm{s}^\mathrm{g}=m \delta N_\mathrm{s}(v_\mathrm{s}=0)$, is intimately connected  with the excess/deficit of particles $\delta N_\mathrm{s}(v_\mathrm{s})$, which according to the general thermodynamic relation is given by $\delta N_\mathrm{s}=-\partial E_\mathrm{s}/ \partial E_\mathrm{F}$. The excess/deficit  of particles for one- ($+$) and quasi-one ($-$) dimensional superfluids is given by
\begin{equation}
\label{ParticleDefficit}
\delta N_\mathrm{s}\approx\pm \frac{\Delta_0}{\lambda \pi E_\mathrm{F}}\sqrt{1-\left(\frac{v_\mathrm{s}}{v_\mathrm{L}}\right)^2}.
\end{equation}
Its absolute value decreases with soliton's velocity $v_\mathrm{s}$ and vanishes at the critical velocity $v_\mathrm{L}$. Note that, it is small in the weak-coupling BCS limit and is not captured by direct counting of the occupied states within the Andreev approximation, that we discuss in Section.~III.

In both one- and quasi-one-dimensional systems, the absolute value of the inertial mass is orders of magnitude larger than the gravitational one, resulting in $\omega_\mathrm{s}/\omega\ll 1$, that makes the soliton motion remarkably slow. Particularly, for the coupling constant $\lambda\approx 0.3$ and the trap period $T=2\pi/\omega\approx 60\,ms$ \cite{Zwierlein1}, we have $|m_\mathrm{s}^\mathrm{g}/m_\mathrm{s}^\mathrm{i}|\approx 10^{-3}$ and the period of soliton oscillations $T_\mathrm{s}\approx 1.9\, s$ is macroscopically large.

Note that the notion of soliton's inertial mass is based on the Taylor expansion of the non-linear soliton spectrum~(\ref{Lorentz}) on $v_s^2$. While an effective mass is indeed a useful, intuitive concept, there is no need for this expansion, as the classical equations of motion for a soliton in a trap can be integrated exactly taking into account the full non-linear energy spectrum~ (\ref{Lorentz}) (which is especially important at high soliton velocities, where the aforementioned approximation breaks down). The corresponding soliton's equation of motion in quasi-one-dimensional superfluid, accounting for the full energy spectrum, is given by
\begin{equation}
\label{EqOfMotion1}
\frac{\ddot{x}_\mathrm{s}}{1-\left(\dot{x}_\mathrm{s}/v_\mathrm{L}\right)^2}+ \frac{\omega_\mathrm{s}^2 x_\mathrm{s}}{1-\left( \omega_\mathrm{s} x_\mathrm{s}/\sqrt{2}v_\mathrm{L}\right)^2}=0. 
\end{equation}
If a soliton is created initially at rest, $v_\mathrm{s}(0)=0$, at a distance $x_\mathrm{s}(0)=x_0$ from the trap center, it is pushed to the trap center and its motion depends only on  a single control parameter $x_0/x_*$, where  $x_*=\sqrt{2}v_\mathrm{L}/\omega_\mathrm{s}$ is the distance from the trap center, at which absolute value of potential energy is equal to the maximal kinetic energy of the soliton $E_\mathrm{s}(0)=2\Delta_0/\pi$. For $x_0\ge x_*$ the initial potential energy is sufficient to accelerate the soliton up to the critical velocity $v_\mathrm{L}$ within one cycle, and the soliton vanishes without reaching the trap center. For $x_0<x_*$ the soliton motion is oscillatory and the equation of motion - Eq.~(\ref{EqOfMotion1}) - can be integrated in terms of elliptic functions as follows
\begin{equation}
\label{EqOfMotion2}
\begin{split}
\frac{\sqrt{2} x_*}{\sqrt{2 x_*^2-x_0^2}}F\left(\theta,\frac{x_0^2}{2 x_*^2-x_0^2}\right) +2 \frac{\sqrt{(2 x_*^2-x_0^2)}}{\sqrt{2 }x_*} \times \\ \left[E\left(\theta,\frac{x_0^2}{2 x_*^2-x_0^2}\right)-F\left(\theta,\frac{x_0^2}{2 x_*^2-x_0^2}\right)\right]=\omega_\mathrm{s} t,
\end{split}
\end{equation}
where $F(\theta,x_0^2/(2 x_*^2-x_0^2))$/$E(\theta,x_0^2/(2 x_*^2-x_0^2))$ is incomplete elliptic integral of the first/second kind, and $\theta=\arccos(x_\mathrm{s}/x_*)$. The time dependencies of soliton's coordinate and velocity, originating from Eq.~(\ref{EqOfMotion2}), are presented in Fig.~4. For $x_0\ll x_*$, the oscillatory motion becomes harmonic, while for $x_0\lesssim x_*$ the non-linearity of the equation of motion, Eq.~(\ref{EqOfMotion1}), becomes important and the soliton trajectory becomes visibly different from simple harmonic. Experimental observation of such anharmonic oscillations can reveal  deviations of soliton's dispersion law from the simple quadratic spectrum $E_\mathrm{s}=m_\mathrm{s}^\mathrm{i}v_\mathrm{s}^2/2$.

\begin{figure}
\label{Fig4}
\begin{center}
\includegraphics[width=7.8 cm]{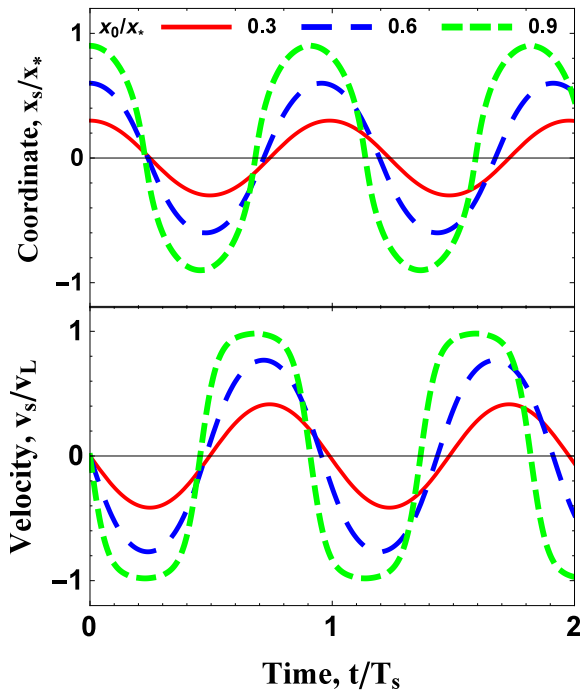}
\caption{(Color online) Shown is dependencies of the soliton's coordinate $x_\mathrm{s}$ and the velocity $v_\mathrm{s}$ on a time $t/T_\mathrm{s}$, where $T_\mathrm{s}=2\pi/\omega_\mathrm{s}$, for different initial positions $x_0/x_*$. For $x_0/x_*\ll 1$ oscillations become harmonic, while for $x_0/x_*\lesssim 1$ the nonlinearity of the equation of motion, given in eq.~(\ref{EqOfMotion1}), becomes important For $x_0>x_*$ soliton achieves the critical velocity $v_\mathrm{L}$ and vanishes without reaching the trap center.}
\end{center}
\end{figure}
\section{Conclusions}
This paper has developed an analytical theory of a moving soliton in a paired fermionic superfluid. The main results are the dependencies of the phase jump across the soliton, its energy and deficit of particles in the core on the soliton velocity. The \emph{only}  approximation used in solving the time-dependent, self-consistent Bogoliubov-de Gennes equations is the Andreev approximation, which involves linearization of the fermion spectrum in the vicinity of the Fermi points. The approximation allows to connect the problem one-to-one to the Gross-Neveu model, for which static solitonic solutions have been studied in detail. We extend the theory to the dynamic situation of a moving soliton. The Andreev approximation is well-justified in the weak-coupling regime $\lambda\ll 1$, and remains reasonable at $\lambda\lesssim 1$ making our extrapolated analytical results of value in that case as well. 

Solitons in fermionic superfluids appear due to a subtle interplay between the  bosonic superconducting order parameter and fermionic quasiparticles. This is contrast to bosonic superfluids, where the Gross-Pitaevskii solitons are structureless. Nevertheless, it was shown that the internal structure of solitons and their physics  evolve smoothly between these regimes across the BEC-BCS crossover. Particularly, solitons in \emph{three-dimensional} fermionic superfluids  were recently investigated numerically in the crossover regime using time-dependent BdG equations~ \cite{SolitonBECBCS1,SolitonBECBCS2,SolitonBECBCS3,SolitonBECBCS4,SolitonBECBCS5,SolitonBECBCS6}. The numerical treatment works in the crossover regime $-1\lesssim (a k_\mathrm{F})^{-1} \lesssim 1$, where $a$ the is the fermion scattering length, but seems to break down in the weak-coupling BCS limit $(a k_\mathrm{F})^{-1}\ll-1$ (where our analytical results are asymptotically exact). This circumstance does not allow us to perform a full comparison between the existing numerical and our analytical results.  However, the velocity dependencies of the soliton profile, energy, phase jump and the deficit of particles, calculated here, are in a good qualitative agreement with the ones obtained numerically on the BCS side of the crossover. Our results can provide a useful reference point for possible future numerical simulations of solitons in this limit.

The internal structure of solitons in a bosonic superfluids differs from their fermionic counterpart, but the two types of solitons have much in common. Particularly, the velocity dependencies of phase jump $2\phi_\mathrm{s}$, energy $E_\mathrm{s}$ and profile of the order parameter $\Psi(z)$ for bosonic superfluid have similar form~\cite{PitaevskiiStringari}

\begin{equation}
\label{GPs}
\begin{split}2\phi_\mathrm{s}=\arccos\left(\frac{v_\mathrm{s}}{c}\right), \quad \; E_\mathrm{s}=\frac{4 \hbar c n_0}{3}\left[1-\left(\frac{v_\mathrm{s}}{c}\right)^2\right]^{\frac{3}{2}},  \\ \frac{\Psi}{\sqrt{n_0}}= \cos(\phi) + \bm{i} \sin(\phi) \tanh \left[ \sin (\phi) \frac{z}{\xi\sqrt{2}}\right],\quad\; \end{split}
\end{equation}

to the  ones in the fermionic case -- see, Eqs.~(\ref{PhaseProfile}), (\ref{Lorentz}) and (\ref{SolitonicProfile}). In Eqs.~(\ref{GPs}), $n_0$ is the equilibrium concentration of bosonic condensate far from the soliton, $\xi$ is its coherence (healing) length, and the critical velocity $c$  is the speed of sound in the bosonic superfluid (in contrast to the fermionic critical velocity, $v_\mathrm{L}$, which is the Landau critical velocity, where the emission of fermionic quasiparticles commences). Also, in contrast to the fermionic superfluid, the notch in the bosonic order parameter $\Psi(z)$ results in an equivalent notch in the particle density. As a result, the Gross-Pitaevskii soliton is accompanied by a macroscopically large deficit of particles
\begin{equation}
\delta N_\mathrm{s}=-\frac{2 \hbar n_0}{mc} \sqrt{1-\left(\frac{v_\mathrm{s}}{c}\right)^2},
\end{equation}
c.f., Eq.~(\ref{ParticleDefficit}). The inertial and gravitational masses of the bosonic soliton are both negative  and their values are connected as $m_\mathrm{s}^\mathrm{i}=2 m_\mathrm{s}^\mathrm{g}$. The soliton oscillation frequency differs from the trap frequency by a factor of $\sqrt{2}$, i.e. $\omega_\mathrm{s}=\omega/\sqrt{2}$. This result is in strong qualitative contrast with the order of magnitude difference between the soliton masses in the BCS fermionic superfluid. There, $\omega_\mathrm{s}\ll\omega$ and the motion of soliton is much slower than that of a bosonic soliton put in the same trap.     


\section*{Acknowledgments}
This research was supported by  DOE-BES DESC0001911 (D.E.),  US-ARO (V.G.), and Simons Foundation. The authors are grateful to Victor Yakovenko, Martin Zwierlein and Lev Pitaevskii for illuminating discussions and a number of useful suggestions.

\bibliography{SolitonBibliorgraphy}

\begin{thebibliography}{43}
\bibitem [{Com()}]{Comment1}%
  \BibitemOpen
  \href@noop {} {}\bibinfo {note} {In a genuine one-dimensional system pairing
  fluctuations are exceptionally strong and destroy long-range coherence. Here
  we take in mind atomic superfluids in cigar-shaped traps, which are
  quasi-one-dimensional systems, for which mean-field theory is resonable
  approximation.}\BibitemShut {Stop}%
\end{thebibliography}

\begin{widetext}

\subsection{Generalized periodic boundary conditions}
Bogoliubov-de Gennes equations (\ref{BdGKSC}) require appropriate boundary conditions. For a uniform superfluid, the simple periodic boundary conditions, $\psi_{\gamma k}^{\alpha}(z+L/2)=\psi^\alpha_{\gamma k} (z-L/2)$ (with $L$ being the system size) apply. 
However, they can not be used in the presence of a soliton, since the order parameter is no longer a periodic function of the coordinate. 
Indeed, while all \emph{local} physical observables [{\em e.g.}, the fermion current $j(z)$, density $\rho(z)$, {\em etc.}] are periodic functions of the coordinate in the closed system [$j(z+L/2)=j(z-L/2)$, $\rho(z+L/2)=\rho(z-L/2)$, {\em etc.}], the order parameter is not periodic, because it has a \emph{global} phase discontinuity across the soliton, and $\Delta(z+L/2)=\Delta(z-L/2)e^{2\bm{i} \phi}$. 

Here, we generalize the simple periodic boundary conditions to the the system with a soliton. The general form of boundary conditions is    
\begin{equation}
\label{BoundaryConditions1}
\psi^\alpha_{\gamma k}(z+L/2) =\hat{B}^\alpha_{\gamma k} \psi^\alpha_{\gamma k}(z-L/2), 
\end{equation}
where $ \hat{B}^\alpha_{\gamma k}(\phi)$ is a matrix (whose explicit form is to be determined) that depends on the phase jump across the soliton. We assume that boundary conditions do not mix states with different quantum numbers and omit the corresponding  indexes $\alpha$,$\gamma$, and $k$, that become redundant. First, we require that the fermion current and density  
\begin{equation}
j(z) =\psi^*(z) \psi(z), \quad \quad  \quad \quad \rho(z) =1+ \psi^*(z) \sigma_z \psi(z) 
\end{equation}
are periodic functions. These conditions lead to the following constrains, $\hat{B}^+\hat{B}=1$ and $\hat{B}^+ \sigma_z \hat{B} =\sigma_z$. The former implies that the matrix $\hat{B}$ is unitary, while the latter allows us to parameterize it by two phases, $\Phi$ and $\Theta$, as follows 
\begin{equation}
\label{BoundaryConditions2}
\hat{B}=e^{i \Phi} \left[\cos (\Theta) + \bm{i} \sin (\Theta) \sigma_z\right].
\end{equation}
Next, assuming the state $\psi(z-L/2)$ to be an eigenvector of the BdG Hamiltonian, $K_\mathrm{BdG}(z-L/2)\psi(z-L/2)=\epsilon \psi(z-L/2)$, we demand that the spatially-translated state, $\psi(z+L/2)$, is an eigenvector of the translated BdG Hamiltonian $K_\mathrm{BdG}(z+L/2)\psi(z+L/2)=\epsilon \psi(z+L/2)$. Note that due to the presence of the phase jump, $\Delta(z+L/2)=\Delta(z-L/2)e^{2\bm{i} \phi}$, the Hamiltonian is not invariant under translation. Using the explicit form of the BdG Hamiltonian (\ref{BdGKSC}), we arrive at 
\begin{equation}
\label{BoundaryConditions3}
\begin{split}
\hat{B}^+ \sigma_x \hat{B}=\sigma_x \cos(\phi) + \sigma_y \sin(\phi), \\
\hat{B}^+ \sigma_y \hat{B}=\sigma_y \cos(\phi) -  \sigma_x \sin(\phi).    
\end{split}
\end{equation} 
The Ansatz (\ref{BoundaryConditions2}) satisfies (\ref{BoundaryConditions3}) if $\Theta=\phi$. Finally, we notice that the superfluid state with the order parameter (\ref{SolitonicProfile}) becomes equivalent to the uniform BCS state at $\phi=0$, since the soliton profile (\ref{SolitonicProfile}) vanishes. Therefore, we must require that $\hat{B}(\phi=0)=\hat{1}$, since $\hat{B}=\hat{1}$ corresponds to the simple periodic boundary conditions. This constraint fixes the remaining parameter $\Phi=0$, and  determines  the unitary matrix $\hat{B}(\phi)$ as follows
\begin{equation}
\label{BoundaryConditions4}
\hat{B}(\phi)=\cos (\phi) + \bm{i} \sin (\phi) \sigma_z.
\end{equation}
The matrix does not depend on the set of indexes $\alpha$, $k$ and $\gamma$ for a continuous Bogoliubov state. 

Let us remark that the boundary condition (\ref{BoundaryConditions4}) can be straightforwardly  generalized to the presence of a soliton train (not relevant here, but of importance to studies of inhomogeneous superconducting states). There, the boundary conditions would have the same form as Eq.~(\ref{BoundaryConditions4}), but with $2\phi$ replaced by the whole phase jump across the train. 

\subsection{Momentum quantization and phase shifts}
\begin{figure}
\label{FigSup}
\begin{center}
\includegraphics[width=8.5 cm]{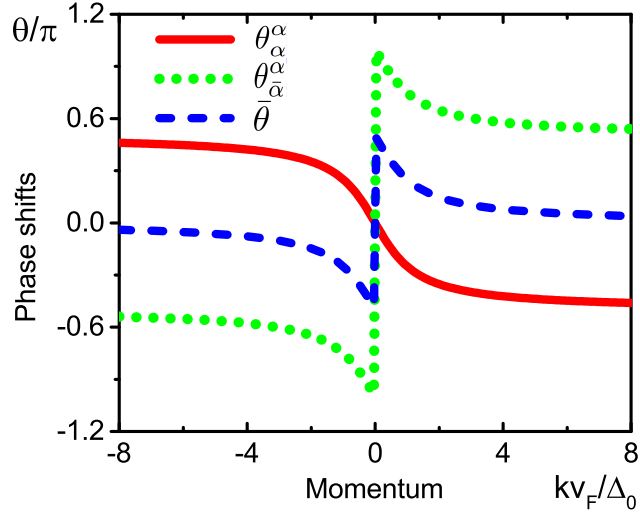}
\caption{(Color online) Phase shifts, $\theta^\alpha_\alpha(k)$, $\theta^\alpha_{\bar{\alpha}}(k)$ and $\bar{\theta}$ [defined in Eqs.~(\ref{PhaseShiftsSup})  and  (\ref{PhaseShiftAverage})], are plotted as a function of momentum (here, the specific value of the  phase jump across soliton is  taken to be $2\phi=\pi$). The dependence remains qualitatively the same  for other values of the phase discontinuity.}    
\end{center}
\end{figure}

The simple periodic boundary conditions, that can be used for a uniform superfluid, determine the standard momentum  quantization rule: $k_n L=2 \pi n$. In the presence of a soliton, momentum quantization is modified and follows from the appropriate boundary conditions (\ref{BoundaryConditions4}). 

Let us rewrite the boundary conditions in terms of  the functions $f^\alpha_{\gamma k, \pm }=u^\alpha_{\gamma k}\pm v^\alpha_{\gamma k}$, as follows
\begin{equation}
\label{BoudaryF}
f^\alpha_{\gamma k, \pm }(z+L/2)=\cos(\phi) f^\alpha_{\gamma k, \pm }(z-L/2)  + \bm{i} \sin(\phi) f^\alpha_{\gamma k, \mp }(z-L/2).
\end{equation}
The functions $f^\alpha_{\gamma k, \pm}$ in the solitonic state are given by
\begin{equation}
\label{WaveFunctionsF}
f^\alpha_{\gamma k, \alpha }(z)=\sqrt{\frac{\epsilon_{\gamma k} + \alpha \Delta_1}{L \epsilon_{\gamma k}}} e^{\bm{i} k z },\quad \quad \quad \quad  f^\alpha_{\gamma k, \bar{\alpha}}(z)= \alpha \sqrt{\frac{\epsilon_{\gamma k }  + \alpha \Delta_1}{L \epsilon_{\gamma k}}} \frac{\hbar v_\mathrm{F}k +\bm{i} \Delta_2(z)}{\epsilon_{\gamma k} + \alpha \Delta_1} e^{\bm{i} k z }.
\end{equation}
Substitution of (\ref{WaveFunctionsF}) into (\ref{BoudaryF}) leads to
\begin{equation}
\begin{split}
[\epsilon_{\gamma k}+ \Delta_0 \cos(\phi)] e^{\bm{i} kL/2}=[\epsilon_{\gamma k}+ \Delta_0 \cos(\phi)]\cos(\phi) e^{-\bm{i} kL/2}  + i [\hbar v_\mathrm{F} k- \bm{i} \Delta_0 \sin(\phi)] \sin(\phi) e^{-\bm{i} kL/2}, \\
[\hbar v_\mathrm{F} k + \bm{i} \Delta_0 \sin(\phi)] e^{\bm{i} kL/2}=[\hbar v_\mathrm{F} k - \bm{i} \Delta_0 \sin(\phi)] \cos(\phi) e^{-\bm{i} kL/2} + \bm{i}  [\epsilon_{\gamma k} + \Delta_0 \cos(\phi)] \sin(\phi) e^{-\bm{i} kL/2},
\end{split}
\end{equation}

for the right Fermi point, and to   
\begin{equation}
\begin{split}
[\epsilon_{\gamma k}- \Delta_0 \cos(\phi)] e^{\bm{i} kL/2}=[\epsilon_{\gamma k}- \Delta_0 \cos(\phi)]\cos(\phi) e^{-\bm{i} kL/2} - \bm{i} [\hbar v_\mathrm{F} k- \bm{i} \Delta_0 \sin(\phi)] \sin(\phi) e^{-\bm{i} kL/2}, \\
[\hbar v_\mathrm{F} k + \bm{i} \Delta_0 \sin(\phi)] e^{\bm{i} kL/2}=[\hbar v_\mathrm{F} k - \bm{i} \Delta_0 \sin(\phi)] \cos(\phi) e^{-\bm{i} kL/2} - \bm{i}  [\epsilon_{\gamma k}- \Delta_0 \cos(\phi)] \sin(\phi) e^{-\bm{i} kL/2},
\end{split}
\end{equation}
for the left Fermi point. \emph{Each pair} of equations can be reduced to $\mathrm{exp}[\bm{i} k L + \bm{i} \theta_\gamma^\alpha (k)]=1$, which yield a momentum quantization rule as follows  $k_n L+ \theta^\alpha_\gamma (k_n)=2 \pi n$. Here $\theta^\alpha_{\gamma}(k)$ is the phase shift, which is given by 
\begin{equation}\label{PhaseShiftsSup}
\theta^\alpha_{\gamma}(k)=\mathrm{arg}\left[\epsilon_{k} \cos(\phi) + \alpha\gamma \Delta_0 -  \bm{i} \alpha \gamma  \hbar v_\mathrm{F} k \sin(\phi) \right].
\end{equation}
The dependence of the phase shifts on momentum is presented in Fig.~S1. Their asymptotic values at infinite momenta are given by 
\begin{equation}
\theta^\alpha_\alpha (\infty)=-\phi, \quad \quad \quad  \theta^\alpha_{\bar{\alpha}} (\infty)=\phi,  \quad \quad \quad 
 \theta^\alpha_\alpha (-\infty)=\phi, \quad \quad \quad  \theta^\alpha_{\bar{\alpha}} (-\infty)=-\phi.
\end{equation}
The number of states split from the left- and right-moving continuous Bogoliubov bands can be calculated with the help of these phase shifts as follows
\begin{equation}
N^\alpha_\alpha= -\int_{-\infty}^\infty \frac{dk}{2\pi} \frac{d \theta^\alpha_\alpha}{dk}=\frac{\phi}{\pi}, \quad \quad \quad 
N^\alpha_{\bar{\alpha}}= -\int_{-\infty}^\infty \frac{dk}{2\pi} \frac{d \theta^\alpha_{\bar{\alpha}}}{dk}=1-\frac{\phi}{\pi}.
\end{equation}
Since there is only one ABS per Fermi point, the total splitting from the continuous bands is equal $N^\alpha_-+N^\alpha_+=1$. The total number of states split from the Bogoliubov states with negative energies is also equal to $N^-_-+N^+_-=1$. 
 
For a calculation of the energy of a superfluid, which is presented in Appendix~C, it is useful to introduce the average phase shift
$\bar{\theta}=(\theta^+_- + \theta^-_-)/2$. Using the the relations
\begin{equation} 
\cos(\theta^\alpha_\gamma)=\frac{\epsilon_k \cos (\phi)+\alpha \gamma \Delta_0}{\epsilon_k+\alpha\gamma \Delta_0\cos(\phi)},  \quad \quad \quad \quad \sin(\theta^\alpha_\gamma)=-\frac{ \alpha\gamma\sin (\phi)}{\epsilon_k+\alpha\gamma \Delta_0\cos(\phi)}
\end{equation}
the average phase shift $\bar{\theta}$ can be calculated as follows 
\begin{equation}
\label{PhaseShiftAverage}
\bar{\theta}=\arctan\left[\sqrt{\frac{1+\cos(\theta^+_-+\theta^-_-)}{1-\cos(\theta^+_-+\theta^-_-)}}\right]=\arctan\left[\frac{\Delta_0 \sin(\phi)}{\hbar v_\mathrm{F}k}\right].
\end{equation}
The dependence of the average phase shift $\bar{\theta} $ on the momentum is presented in Fig.~5.

\subsection{Calculation of the soliton energy in the co-moving and laboratory  frames}

The energies of a fermionic superfluid in the co-moving ($E^{\mathrm{K}}$) and laboratory ($E^{\mathrm{H}}$) frames can be determined from the Hamiltonians $K_\mathrm{BdG}$ [defined in Eq.~ (\ref{BdGKSC})] and $H_\mathrm{BdG}$ [defined in Eq.~ (\ref{BdGH})], respectively. The energies of Bogoliubov states of $K_\mathrm{BdG}$ and $H_\mathrm{BdG}$ differ by the shift $\delta \epsilon^\alpha=\alpha v_\mathrm{s} p_\mathrm{F}$, while the occupation numbers are the same and correspond to $K_\mathrm{BdG}$, since in the co-moving frame the solitonic texture is time-independent and the superfluid achieves thermal equilibrium. The difference in energy between a superfluid  with a soliton and the uniform BCS state can be presented as the sum $$
E^\mathrm{K(H)}=E_\mathrm{\Delta}+E_{\mathrm{c}}^\mathrm{K(H)}+E_{\mathrm{ABS}}^\mathrm{K(H)}.
$$
The first term, $E_\mathrm{\Delta}$, in this equation comes directly from the non-uniformity of the order parameter, it does not depend on the energy shift, and is given by
\begin{equation}
\label{EnergyDelta}
E_\mathrm{\Delta}= \int dz \frac{(|\Delta|^2-\Delta_0^2)}{V} = - \sum_{\vec{k}} \frac{2 \hbar v_\mathrm{F} \Delta_0 \sin{\phi}}{\epsilon_k},
\end{equation}
where we have eliminated the coupling constant $V$ using the self-consistency equation (\ref{SelfConsistent}) for the uniform BCS state. Contributions $E_{\mathrm{c}}^\mathrm{K}$ and $E_{\mathrm{c}}^\mathrm{H}$ originate from filled continuous Bogoliubov states, whose occupations are not influenced by the energy shift. Therefore, they can be calculated with the help of phase shifts (\ref{pshift}) as follows
\begin{equation}
E_\mathrm{c}^\mathrm{H}=\sum_{\alpha}\left[{N^{\alpha}_{-}} \Delta_0 + \sum_{k} \theta^{\alpha}_{-} \frac{\partial \epsilon_{k}}{\partial k}\right] \quad \quad \quad \quad  E_\mathrm{c}^\mathrm{K}=\sum_{\alpha}\left[{N^{\alpha}_{-}} \Delta_0 +  \sum_{k} \theta^{\alpha}_{-} \frac{\partial \epsilon_{k}}{\partial k}\right] - v_\mathrm{s}p_\mathrm{F} (N^+_- - N^-_-).   
\end{equation}
The last term in $E^\mathrm{K}_\mathrm{c}$ originates from a difference in the number of states split from the right- and the left-moving filled bands. The energy $E_\mathrm{c}^\mathrm{H}$ can be calculated as follows 
\begin{equation}
E_\mathrm{c}^\mathrm{H}=\Delta_0 + 2 \int_{0}^\infty \frac{dk}{\pi} \bar{\theta}(k) \frac{d \epsilon_k}{d k}=\frac{2 \Delta_0 \sin(\phi)}{\pi} - 2 \int_{0}^\infty \frac{dk}{\pi} \frac{\bar{\theta}(k)}{dk} \epsilon_k=\frac{2 \Delta_0 \sin(\phi)}{\pi} +   \int_0^\infty \frac{dk}{\pi} \frac{ 2 \hbar v_\mathrm{F} \Delta_0 \epsilon_k \sin(\phi)}{ (\hbar v_\mathrm{F} k)^2 +[\Delta_0 \sin(\phi)]^2} .  
\end{equation}
Here, we have taken into account that the total number of states split from the Bogoliubov hole bands for the right and left Fermi points is $N^+_- + N^-_-=1$ and introduced the average phase shift $\bar{\theta}=(\theta^+_- + \theta^-_-)/2=\arctan[\Delta_0 \sin(\phi)/\hbar v_\mathrm{F}k]$, calculated in Appendix~B. Combining with (\ref{EnergyDelta}) and performing an integration, we arrive at 
\begin{equation}
E_\mathrm{\Delta}+E_\mathrm{c}^\mathrm{H}= \frac{2\Delta_0}{\pi} \left[ \sin(\phi) + \left( \frac{\pi}{2} -\phi\right) \cos(\phi) \right], \quad \quad \quad \quad  E_\mathrm{\Delta}+E_\mathrm{c}^\mathrm{K}= E_\mathrm{\Delta}+E_\mathrm{c}^\mathrm{H} - v_\mathrm{s} p_\mathrm{F} \left(1 -\frac{2 \phi}{\pi} \right). 
\end{equation}
The last contributions $E^\mathrm{H}_\mathrm{ABS}$ and $E^\mathrm{K}_\mathrm{ABS}$ originate from the ABS. Both energies and occupations of ABS are influenced by the energy shift, $\delta \epsilon=\alpha v_\mathrm{s} p_\mathrm{F}$. Hence, it is instructive to consider them separately.  In the co-moving frame,  the energy is given by $E_{\mathrm{ABS}}^\mathrm{K}=-[v_\mathrm{s}p_\mathrm{F} - \Delta_0 \cos(\phi)]  \tanh\{ [v_\mathrm{s}p_\mathrm{F} - \Delta_0 \cos(\phi)]/T\}$. The zero-temperature limit $T\ll |v_\mathrm{s}p_\mathrm{F} - \Delta_0 \cos(\phi)|$ is well-defined and the energy at $T=0$ is given by $E_{\mathrm{ABS}}^\mathrm{K}=-|v_\mathrm{s}p_\mathrm{F} - \Delta_0 \cos(\phi)|$. Combining all contributions together, we get the energy of a superfluid with a soliton in the co-moving frame to be  
\begin{equation}
\label{EnergyKSup}
E^\mathrm{K}(\phi,v_\mathrm{s})=\frac{2 \Delta_0}{\pi} \left[ \sin(\phi)+\left( \frac{\pi}{2} - \phi\right)\cos(\phi)\right] - v_\mathrm{s} p_\mathrm{F} \left(1-\frac{2 \phi}{\pi}\right)
- |v_\mathrm{s}p_\mathrm{F} - \Delta_0 \cos(\phi)|.
\end{equation}
In the laboratory frame, the contribution of ABS is given by $E_{\mathrm{ABS}}^\mathrm{H}= \Delta_0 \cos(\phi)  \tanh\{[v_\mathrm{s}p_\mathrm{F} - \Delta_0 \cos(\phi)]/T\}$. In the zero temperature limit, it tends to $E_{\mathrm{ABS}}^\mathrm{H}= -\Delta_0 \cos(\phi)  \Theta_\mathrm{H}[\Delta_0 \cos(\phi)-v_\mathrm{s}p_\mathrm{F}]$ and the energy of the superfluid in the laboratory frame is given by 
\begin{equation}
\label{EnergyHSup}
E^\mathrm{H}(\phi,v_\mathrm{s})=\frac{2 \Delta_0}{\pi} \left[ \sin(\phi)+\left( \frac{\pi}{2} - \phi\right)\cos(\phi)\right] 
 -\Delta_0 \cos(\phi)  \Theta_\mathrm{H}\left[\Delta_0 \cos(\phi)-v_\mathrm{s}p_\mathrm{F}\right],
\end{equation}
where $\Theta_\mathrm{H}$ is Heaviside step function. However, in this case, the zero temperature limit is \emph{ill-defined} since $E_{\mathrm{ABS}}^\mathrm{H}(\phi,v_\mathrm{s})$ [and hence $E^\mathrm{H}(\phi,v_\mathrm{s})$ too] is not a smooth function of its arguments. The energy has a jump across the line $\Delta_0 \cos(\phi)-v_\mathrm{s}p_\mathrm{F}=0$, which corresponds to the solitonic profile (\ref{PhaseProfile}). Hence the calculation of the energy of a superfluid in the solitonic state, which has the phase profile (\ref{PhaseProfile}) requires a more delicate approach. In the solitonic state, both energies $\epsilon_{\mathrm{ABS},\mathrm{s}}^\alpha=-\alpha v_\mathrm{s} p_\mathrm{F}$ and occupations of ABS adjust to soliton's motion. Hence the contribution of ABS is well defined and is given by 
\begin{equation}
E^\mathrm{H}_\mathrm{ABS}= \epsilon^+_\mathrm{ABS,s} n^+_\mathrm{ABS,s} + \epsilon^-_\mathrm{ABS,s} n^-_\mathrm{ABS,s}=\frac{ v_\mathrm{s} p_\mathrm{F}}{\pi}\left[2 \arccos \left(\frac{v_\mathrm{s}}{v_\mathrm{L}} \right) -\pi\right],
\end{equation}
where $v_\mathrm{L}=\Delta_0/p_\mathrm{F}$ is the critical velocity within the Landau criterion. Collecting all other contributions, $E_\mathrm{\Delta}(\phi_\mathrm{s}(v_\mathrm{s}))$ and $E^\mathrm{H}_\mathrm{c}(\phi_\mathrm{s}(v_\mathrm{s}),v_\mathrm{s})$, we obtain the energy of the soliton in the laboratory frame as follows 
\begin{equation}
E_\mathrm{s}(v_\mathrm{s})=E^\mathrm{H}(\phi_\mathrm{s}(v_\mathrm{s}),v_\mathrm{s})=\frac{2 \Delta_0}{\pi}\sqrt{1-\left(\frac{v_\mathrm{s}}{v_\mathrm{L}}\right)^2}.
\label{Lorentz2}
\end{equation}

\end{widetext}

\end{document}